\documentclass[a4paper,twoside]{article}
\usepackage{qic,epsfig}

\usepackage{amsmath}
\usepackage{cite}
\usepackage{MissingT}

\renewcommand{\includegraphics}[2][\relax]
             { \ifx\relax#1
                 \epsfig{file=#2.eps}
               \else
                 \epsfig{file=#2.eps,#1}
               \fi
             }
\renewcommand{\caption}[1]{\fcaption{#1}}

\begin{document}
\setlength{\textheight}{8.0truein}    

\runninghead{Threshold error rates for the toric and surface codes}
            {D. S. Wang, A. G. Fowler, A. M. Stephens, L. C. L. Hollenberg}

\normalsize\textlineskip
\thispagestyle{empty}
\vspace*{0.88truein}
\setcounter{page}{1}
\alphfootnote
\fpage{1}

\centerline{\bf THRESHOLD ERROR RATES FOR THE TORIC AND SURFACE CODES}
\vspace*{0.37truein}
\centerline{\footnotesize
        D. S. WANG\footnote{dswang@physics.unimelb.edu.au} ,
        A. G. FOWLER, A. M. STEPHENS, L. C. L. HOLLENBERG}
\vspace*{0.015truein}
\centerline{\footnotesize\it Centre for Quantum Computer Technology,}
\baselineskip=10pt
\centerline{\footnotesize\it School of Physics, University of Melbourne,}
\baselineskip=10pt
\centerline{\footnotesize\it Victoria 3010, Australia}
\vspace*{0.21truein}

\begin{abstract}
The surface code scheme for quantum computation features a 2d array of
nearest-neighbor coupled qubits yet claims a threshold error rate
approaching $1\%$ \cite{Raus07d}.  This result was obtained for the
toric code, from which the surface code is derived, and surpasses all
other known codes restricted to 2d nearest-neighbor architectures by
several orders of magnitude.
%
%
We describe in detail an error correction procedure for the toric and
surface codes, which is based on polynomial-time graph matching
techniques and is efficiently implementable as the classical
feed-forward processing step in a real quantum computer.  By direct
simulation of this error correction scheme, we determine the threshold
error rates for the two codes (differing only in their boundary
conditions) for both ideal and non-ideal syndrome extraction
scenarios.  We verify that the toric code has an asymptotic threshold
of $p_{\textrm{th}} = 15.5\%$ under ideal syndrome extraction, and
$p_{\textrm{th}} = 7.8 \times 10^{-3}$ for the non-ideal case, in
agreement with \cite{Raus07d}.  Simulations of the surface code
indicate that the threshold is close to that of the toric code.
\end{abstract}

\vspace*{10pt}
\vspace*{3pt}
\vspace*{1pt}\textlineskip


\section{Introduction}

Quantum computation is the manipulation of quantum information,
typically in the form of \emph{qubits}, the quantum analogue of the
classical bit \cite{Niel00}.  Qubits differ fundamentally from their
classical counterparts as they can be placed in arbitrary
superpositions of their basis states: $\alpha \ket{0} + \beta
\ket{1}$.  This freedom allows for novel new prospects in computation,
as demonstrated by the existence of quantum algorithms outperforming
existing classical algorithms \cite{Shor94b, Grov96}.
Quantum algorithms must necessarily make use of superposition states
and entanglement to be distinguishable from, and potentially
outperform, classical algorithms.  Thus the preservation of the
quantum state is crucial.  Interactions with the environment
(decoherence) will inevitably occur, altering the fragile quantum
state thus leading to unreliable output.  Quantum error correction
\cite{Shor95, Cald95, Stea96} may be employed to counteract this.

Error correction makes use of redundant information in order to
correct for physical errors; the higher the redundancy, the more
locations for errors and yet simultaneously the more errors that can
be corrected.  The balance between increased errors and corrective
ability gives rise to a threshold error rate \cite{Knil96b}, below
which error correction reduces the effective error rate.  If all
physical gates are constructed with a failure rate below this
threshold error rate, quantum error correction enables arbitrary
length quantum computation to be achieved.
In the case of the surface codes, logical qubits are created from
pairs of holes in a lattice of qubits.  One logical operation is a
chain of physical operations connecting together pairs of holes, the
other is a ring encompassing one hole.  The minimum number of errors
required to cause logical failure can be increased by increasing the
separation between holes and the circumference around holes, thus the
logical error rate is exponentially suppressed assuming random
independent errors.

Threshold error rates are highly dependent on the underlying
assumptions.  For example, given long-range interactions and a very
large number of qubits, one finds the threshold for fault-tolerant
computation to be over $3\%$ \cite{Knil04c}.  At the other end of the
scale, if one assumes the most restrictive arrangement, a linear
nearest-neighbor architecture, the threshold is approximately
$10^{-5}$ \cite{stephens2009atc}.

More recently, the toric code \cite{Kita97b} has kindled interest.
The original scheme arranges a 2d array of nearest-neighbor coupled
qubits on the surface of a torus which may be mapped onto a simple 2d
plane with periodic boundary conditions (figure \ref{fig:toric code
lattice}a).  Computation on the toric code has since been extended to
the surface code which foregoes the need for periodic boundary
conditions \cite{Raus07, Raus07d, Fowl08}, relaxing the demands
imposed on physical realization.  In this paper, we will distinguish
the \emph{toric code} as originally conceived by Kitaev from this
extended 2d \emph{surface code} with its hard boundaries.

The toric code threshold computed in \cite{Raus07d} was surprisingly
high, $p_{\mathrm{th}} = 7.5 \times 10^{-3}$.  In comparison, given
the same 2d nearest-neighbor physical architecture, the concatenated
$7$-qubit Steane code has a threshold of $1.85 \times 10^{-5}$
\cite{Svor06}, while the concatenated Bacon-Shor code has a threshold
of $2.02 \times 10^{-5}$ \cite{spedalieri2008llt}.  In order to verify
the toric code threshold, we simulate the average time a logical qubit
encoded in either code retains its logical state using error
correction procedures one can implement in practice.  Note that
classical computation is required to diagnose the errors from the
quantum circuits in a real life quantum computer, and that the error
correction methodology applied in our simulations is applicable
without modification to a real quantum computer.

Since the two codes differ only in their boundary conditions, one
expects that the asymptotic threshold for the surface code to be
identical to the toric code but this has yet to be numerically
demonstrated.  We justify this claim for when the syndrome extraction
circuits are error-free (\emph{ideal}) and error-prone
(\emph{non-ideal}).  We show that the toric code has an asymptotic
threshold of $p_{\mathrm{th}} = 15.5\%$ under ideal syndrome
extraction, and $p_{\mathrm{th}} = 7.8 \times 10^{-3}$ for the
non-ideal case, coinciding with \cite{Raus07d}.  In both cases, the
surface code threshold lies in the vicinity of this value.

This paper is organized as follows.  Section \ref{sec:the toric code}
briefly reviews the toric code and the surface code.  Section
\ref{sec:syndrome extraction} presents the syndrome extraction
circuits used throughout the simulations.  An efficient method to
correct errors is described in section \ref{sec:error correction}.  We
also discuss several optimizations rendering the method more
practical.  Section \ref{sec:time to failure simulation} describes the
simulation details, and results from the toric code and the surface
code simulations are contrasted.


\section{The toric and surface codes}
\label{sec:the toric code}

\begin{figure}
\centering
(a) \includegraphics[width=0.4\linewidth]{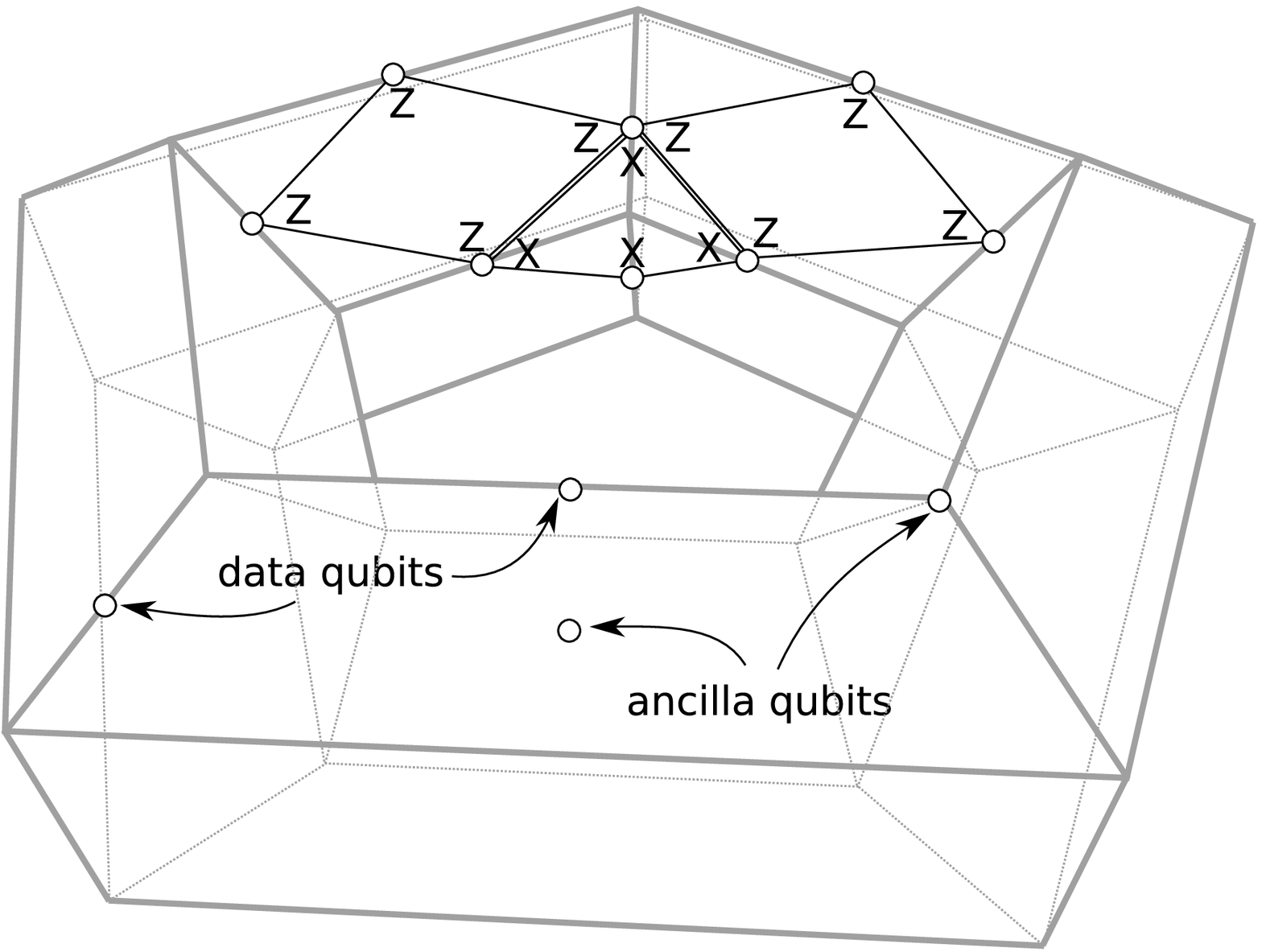}
(b) \includegraphics[width=0.3\linewidth]{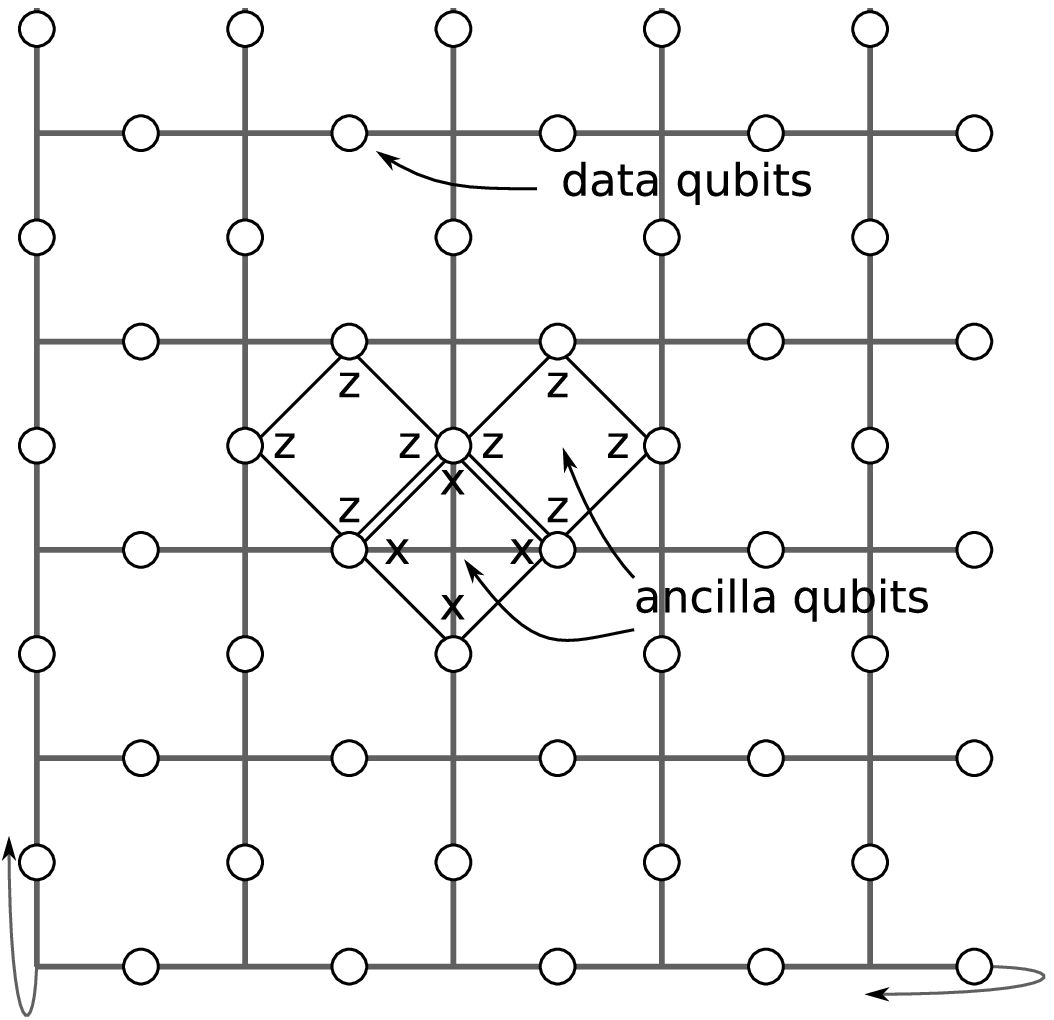}

\caption{%
(a) Toric code qubit arrangement shown in 3d.
(b) Equivalent lattice represented in 2d with periodic boundaries.
Stabilizer generators are the tensor products of the Pauli-$Z$ and
Pauli-$X$ matricies around faces and intersections respectively.
}
\label{fig:toric code lattice}
\end{figure}

We introduce the essential elements of the toric and surface codes
following the ideas of \cite{bravyi2001qcl}.
Consider a collection of qubits arranged on the surface of a torus as
shown in figure \ref{fig:toric code lattice}a, or more easily
represented as a regular 2d array of qubits with periodic boundary
conditions (figure \ref{fig:toric code lattice}b).  The qubits are
divided into two categories: data qubits located on the lines of the
grid, and ancilla qubits on the faces and the intersections.

The stabilizer generators for the toric code are the tensor products
of the Pauli-$Z$ matrices, $Z$, on the four data qubits around each
face, and the tensor products of the Pauli-$X$ matrices, $X$, on the
four qubits around each intersection.  Neighboring stabilizers share
two data qubits ensuring that the $X$ and $Z$-stabilizers commute.
Using the ancilla qubits, the eigenvalues of these stabilizers may be
measured whilst still preserving a quantum state.  We will always
assume that the computer is initialized to the simultaneous $+1$
eigenstate of every stabilizer.

\begin{figure}
\centering
\includegraphics[width=0.4\linewidth]{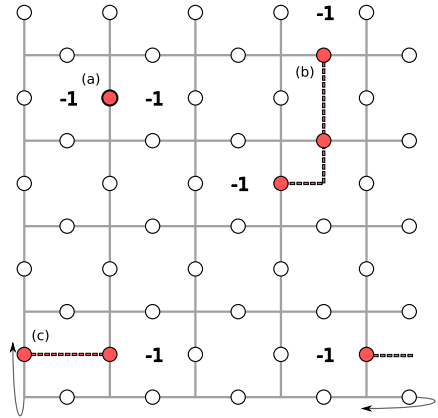}

\caption{The state is initialized to the $+1$ eigenstate of all
stabilizers.  Shaded qubits indicate locations of $X$-errors.
(a) A single $X$-error toggles the eigenvalue measured in the adjacent
faces in the next syndrome extraction cycle.  $Z$-errors are similarly
detected by the neighboring intersections.
(b) When errors are closely spaced, one observes only the terminals of
a continuous chain.
(c) Error chains can wrap around the boundaries in the toric code, but
not in the surface code.
}
\label{fig:toric code example}
\end{figure}

An $X$-error on a data qubit anti-commutes with the two adjacent
$Z$-stabilizers.  Assuming no others errors occur, we observe a change
in the measured eigenvalue of the adjacent $Z$-stabilizers, from $+1$
to $-1$.  Similarly, $Z$-errors result in changes in the measured
eigenvalues of two adjacent $X$-stabilizers.  In general, if many
closely separated errors occur, one observes the \emph{terminals} of
chains of errors (figure \ref{fig:toric code example}).


The interspersed ancilla allow for these eigenvalues to be measured
using only local interactions, via the circuits shown in figure
\ref{fig:syndrome circuits}.  The configuration of eigenvalues
measured by the ancilla on the faces forms the $X$-syndrome from which
$X$-errors may be corrected.  Similarly, $Z$-errors are corrected
using syndrome information from ancilla located on the intersections.
The details of how the syndrome is used to correct errors are
presented in section \ref{sec:error correction}.

\begin{figure}
\centering
\includegraphics[width=0.6\linewidth]{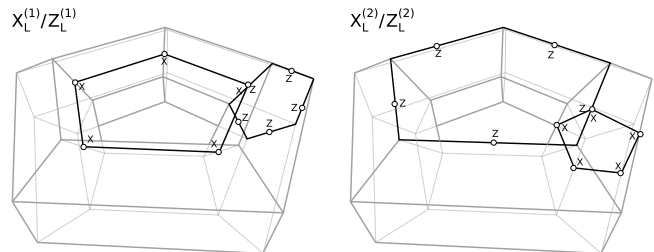}

\caption{Two logical-$X$ and logical-$Z$ operators in the toric code.
Other rings of single-qubit $X$ or $Z$ operators on the lattice can
either be deformed to products of the above rings, or contracted down
to a single stabilizer generator.}
\label{fig:toric code logical operations}
\end{figure}

Logical operations can be associated with the four
\emph{non-homotopic} closed paths around the torus which cannot be
contracted to a single stabilizer generator, shown in figure
\ref{fig:toric code logical operations}.  (Two paths are
\emph{homotopic} if one can be continuously deformed into the other,
in this case by multiplying by stabilizers).  This particular topology
defines two logical qubits: one from $X_L^{(1)}$ and $Z_L^{(1)}$, the
other from $X_L^{(2)}$ and $Z_L^{(2)}$.  These closed rings always
overlap at an odd number of sites, yielding the correct commutation
relations between logical operations.  The \emph{distance} of a code,
$d$, is the length of the shortest logical operation.

\begin{figure}
\centering
\includegraphics[width=0.6\linewidth]{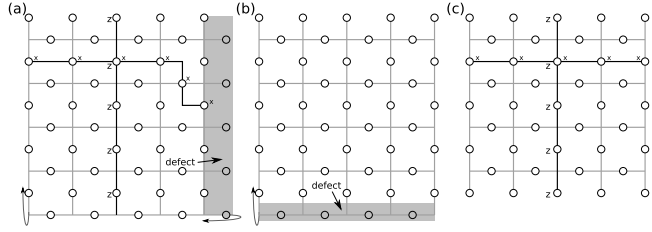}

\caption{%
(a) A smooth periodic defect is introduced into the toric code,
eliminating one periodic boundary.  The logical-$X$ operation now
becomes any chain of $X$ operations connecting the left and right
boundaries together.
(b) A rough periodic defect is introduced to replace the remaining
periodic boundary with a hard boundary.
(c) The 2d surface code and its logical operations.}
\label{fig:toric code to surface code}
\end{figure}

In the absence of long-range physical gates, this topological code is
difficult to realize as one would need to create a quasi-3d quantum
computer.  The surface code is a variation on the toric code, whereby
the periodicity is removed \cite{bravyi2001qcl}.  The construction is
most easily seen as the introduction of two \emph{defects}: regions in
the torus where ancilla qubits are no longer measured.  Data qubits
within the defect are no longer required and the stabilizers must be
changed to reflect this.  This process is illustrated in figure
\ref{fig:toric code to surface code}.

The first ``smooth'' defect cuts the torus into a cylinder.  A smooth
defect is a contiguous region of $Z$-stabilizer generators which we
choose to ignore.  In doing so, one eliminates a column of qubits and
the horizontal periodicity.  In addition, the $Z_L^{(2)}$ operation is
lost, and the logical-$X$ operation becomes any chain of $X$
operations connecting the left and right boundaries together.
Similarly, the vertical periodicity can be removed by introducing a
``rough'' defect --- a defect associated with a contiguous region of
$X$-stabilizer generators.  This defect removes the $X_L^{(2)}$, so
that there still remains one degree of freedom, defining a logical
qubit.  The end result is a code implementable on a regular 2d array
of qubits with only local interactions.
The construction of the surface code shown here will suffice for this
paper.  More detailed accounts can be found in \cite{agarwal1998mss,
freedman1998ppa}.

\section{Syndrome extraction}
\label{sec:syndrome extraction}

\begin{figure}[h]
\centering
\includegraphics{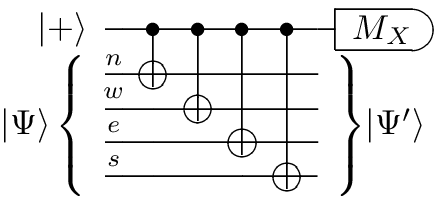}
\includegraphics{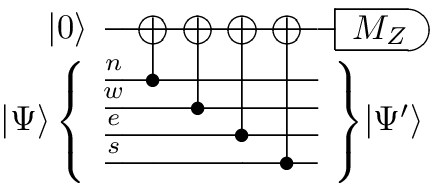}

\caption{Circuits used to measure $X$-stabilizers (left) and
$Z$-stabilizers (right).}
\label{fig:syndrome circuits}
\end{figure}

Syndrome extraction on the toric code is done simultaneously for both
$X$ and $Z$ syndromes.  In our simulations, this is achieved using the
circuits in figure \ref{fig:syndrome circuits}.
First, the ancilla qubits are prepared in their designated states.
Then each ancilla interacts with the data qubits to its north, west,
east and south (in that order).  Finally, the ancilla are measured in
the designated bases.  This particular interaction order leads to
ancilla qubits being in a simple product state at the end of the
cycle, as each syndrome circuit can be shown to occur strictly before
or after its neighboring syndrome circuits.

\begin{figure}
\centering
\includegraphics[width=0.6\linewidth]{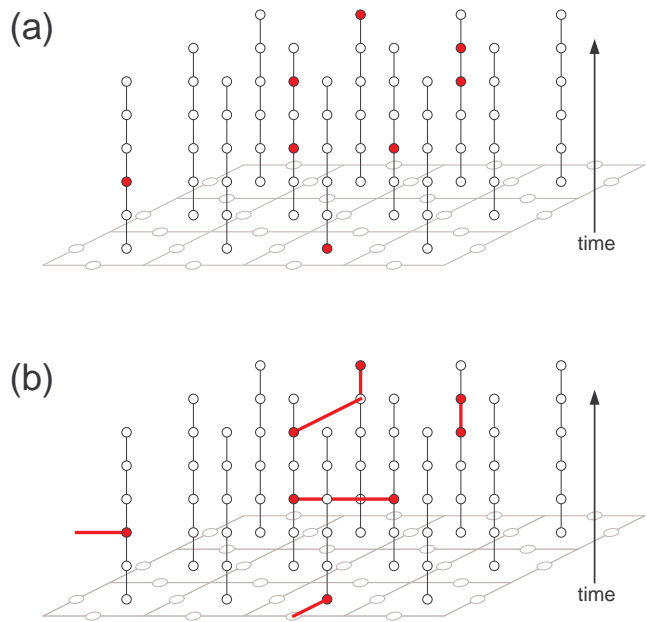}

\caption{%
(a) colored dots indicate the locations in space and time where the
reported syndrome is different from that in the previous timestep.
(b) Changes in syndrome mark the terminals of chains of errors, thus
matching identifies a set of corrections to apply to correct the code.
}
\label{fig:faulty syndrome}
\end{figure}

Since syndrome extraction is itself a physical process, it too is
prone to errors.  Under faulty syndrome extraction, error correction
works in these schemes by collating the eigenvalue measurements over
the duration of the computation, forming a 3d syndrome structure
(figure \ref{fig:faulty syndrome}).  The syndrome is now the
\emph{change} in eigenvalues measured between sequential timeslices,
just as the syndrome for ideal syndrome extraction was the change from
a $+1$ eigenvalue to a $-1$ eigenvalue.  Error correction remains the
task of finding the most likely set of errors that is consistent with
the observed syndrome, which now is in $3$-dimensions: two spatial and
one temporal.

\section{Error correction}
\label{sec:error correction}

The syndrome extracted gives us enough information to correct the
state back to a codeword (ie. the simultaneous $+1$ eigenstate of all
stabilizers), however we have yet to describe a method for doing this.
Further, the method should be \emph{efficient} (ie. runs in
polynomial-time in the number of physical errors) for a quantum
computer implementing this code to outperform classical computation.
This condition prevents us from performing an exhaustive search to
generate the observed syndrome as the search space grows exponentially
with the lattice size.

One efficient method to error correct is by \emph{matching}.  A
perfect matching on a graph is a subgraph in which every node in the
graph has exactly one edge.  Polynomial-time minimum-weight perfect
matching algorithms exist based on Edmonds' blossom algorithm
\cite{edmonds1965pta, cook1997cmw}.

We will first focus on the toric code, assuming ideal (error-free)
syndrome extraction.  The problem is first formed as a graph.  The
observed eigenvalue changes form the nodes on a graph.  Each pair of
nodes is connected by a weighted edge, which is associated with a
chain of errors producing those two observed terminals.  A perfect
matching of the graph is then a set of error chains reproducing the
syndrome.
As there are many possible error chains linking any two nodes, we
choose any of the possible chains of maximum probability.  In our
simulations, since all sites have equal probability of error, this
equates to the shortest length error chains.  An edge linking two
nodes is assigned a weight equal to this shortest length.  Thus a
minimum-weight perfect matching reproduces the observed syndrome using
the fewest errors.  Since each edge in the matching represents one
error chain, in order to error correct one applies corrections on the
qubits along the error chains given by the matching.

\begin{figure}
\centering
\includegraphics[width=1.0\linewidth]{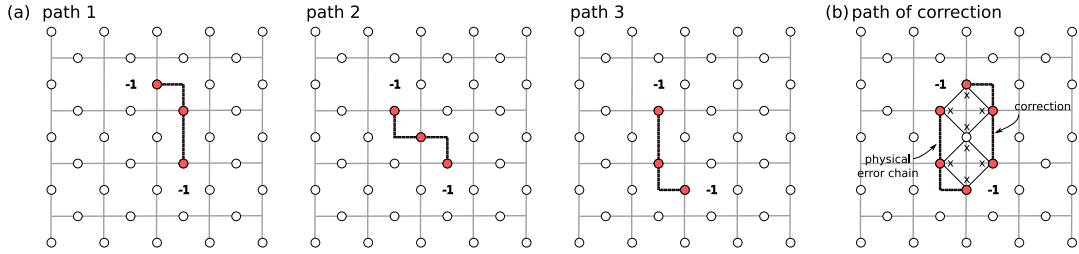}

\caption{%
(a) Three equally weighted error chains with identical syndromes.
(b) The syndrome observed is due to physical errors on the data qubits
along path $3$.  Unable to distinguish between the three options, we
arbitrarily choose to correct along path $1$.  Since the resultant bit
flips (including the physical errors) are a product of stabilizer
generators, the original logical state is restored.
}
\label{fig:degenerate syndromes}
\end{figure}

Consider figure \ref{fig:degenerate syndromes}a, showing three
different physical error chains producing identical syndromes.  When
such ambiguities arise, there is no way to distinguish which
correction to apply.  Suppose that we choose to always correct the
qubits along path 1, but the physical errors occured along path 3.
Under ideal syndrome extraction, it can be shown that for fewer than
$\floor{(d+1)/2}$ errors, any chain connecting the syndrome changes is
homotopic to the physical error chain.  In other words, such
corrections will always form rings of errors, which are products of
the stabilizer generators (figure \ref{fig:degenerate syndromes}b).
Thus the logical state is restored.

Instead of each edge representing only one error chain, one could sum
over all the different ways of generating the observed terminals and
weigh it accordingly.  While this has not been done, it could
potentially improve the error correction procedure.

Generalising this method to incorporate errors during syndrome
extraction is straightforward.  Recall that it is the changes in
eigenvalue measurements at the same location between sequential
timeslices which contribute to the syndrome.  These observed changes
still mark the terminals of error chains, thus we can use the same
formalism.  They form the nodes in a graph, and weighted edges connect
pairs of nodes.

A segment of an error chain spanning through time denotes an incorrect
syndrome measurement, for example due to a preparation or a
measurement error in the ancilla qubit.  In practice, such errors may
occur at different rates compared to data errors.  One could then
weigh time-separated segments of the error-chains differently to
equivalent length space-separated error-chain segments in order to
improve the performance of error correction.  The ideal syndrome
extraction limit is the specialised case where time-separated segments
bear infinite weight, thereby severing nodes in one timeslice from
nodes in other timeslices.  In our simulations, we will take the
simplistic option where space-separated and time-separated segments
are weighted equally for a pessimistic threshold.

During the final step of an algorithm, syndrome information is
extracted by error free circuits so that the final syndrome
measurement can always be assumed to be correct.  Coupled with perfect
initialisation, there will always be an even number of nodes ensuring
a perfect matching is always possible.

\begin{figure}
\centering
\includegraphics{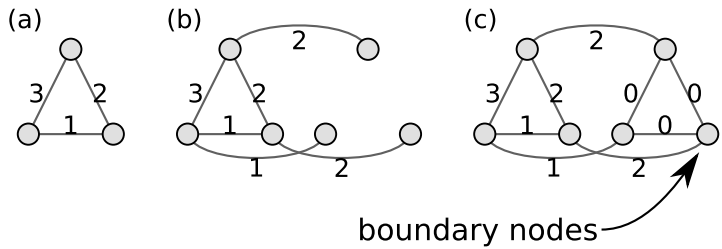}

\caption{
(a) Graph as formed by toric code formulation does not account for
boundaries.
(b) Boundaries included, but matching this graph does not correct
errors.
(c) Final graph which, once matched, returns the desired correction.
}
\label{fig:surface code graph}
\end{figure}

The current method must be adapted to the surface code due to its
non-periodic boundaries; an error on the boundary or a chain of errors
leading to the boundary will generate only one terminal whence no
perfect matching is possible.  The extension is illustrated in figure
\ref{fig:surface code graph}.  First, we determine the closest
boundary to each terminal.  For each node in the original graph, we
create a unique boundary node.  An edge between a node and its
boundary represents an error chain stemming from the closest boundary,
and is weighted accordingly.  If our extension were stopped here, one
would quickly observe that the only perfect matching possible is each
node connected to its own boundary, which has dire consequences.  To
remedy this, all boundary nodes are inter-connected by weight-$0$
edges, so that when two nodes are matched, their respective boundaries
can also be matched with no penalty.

Though we have only discussed space-like boundaries, in practice we
cannot extract the final syndrome exactly.  One can take this into
account by introducing time-like boundaries.  Note that there is still
only one boundary node per terminal; the edge connecting a node to its
boundary now takes the lower weighted choice between space and time
boundaries.

\begin{figure}
\centering
\includegraphics{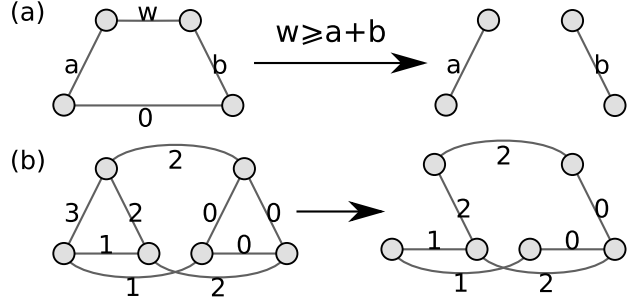}

\caption{
(a) Optimization condition.  Two nodes are connected to their
boundaries by weight-$a$ and weight-$b$ edges respectively.  If the
edge joining the two nodes together has weight $w \geq a + b$, it can
be safely discarded without affecting the minimum-weight matching.
(b) Optimization applied to the graph of figure \ref{fig:surface code
graph}c.
}
\label{fig:optimization}
\end{figure}

The modification of the graph due to the boundaries in this manner
gives rise to interesting optimizations.  First and foremost, it
permits the exclusion of highly weighted edges without fear of loss of
the minimum-weight perfect matching (figure \ref{fig:optimization});
only when it is less costly to join two nodes together (weight $w$)
than each to their respective boundaries (weight $a+b$) must the
former be considered.  This condition discards only edges that we know
from the geometry of the situation cannot be in the minimum-weight
matching, allowing one to match a less highly connected graph than
initially constructed.  The case where $w=a+b$ is ambiguous and
applying the optimization means we always prefer to match the nodes
individually to their boundaries than to other nodes when the
probabilities are equal.  The optimization itself does not affect the
weight-sum of the minimum-weight matching, hence it is \emph{not} an
approximation.

This optimization is of particular interest in the low physical error
rate limit, where the terminals are so sparse that the graph breaks
down into independent components.  Each graph component can be matched
independently.  This in effect simplifies the global optimization
problem to several local problems.  The independence of graph
components ensures the task can be easily parallelized, though we have
not taken this direction.

Furthermore, the time separation between the latest node in a graph
and the current time ensures that there exists a minimum edge weight
connecting any node in that component to any future node.  Once this
minimum edge weight exceeds twice the distance to the boundary, the
optimization condition will always be satisfied.  The graph component
becomes disconnected from all others; no more nodes or edges need to
be added to it.  Hence it can be matched, even as the quantum
computation is running.


\section{Time to failure simulation}
\label{sec:time to failure simulation}

Thus far we have discussed how syndrome information is extracted from
the toric code and the surface code, and described an efficient
implementation of the error correction procedure under both of these
schemes.  This forms the basis for our threshold error rate
simulations.

The threshold error rate is derived from four processes:
initialisation, readout, single-qubit gates (memory) and two-qubit
gates (controlled-not).  Each of these processes have an associated
error rate and duration.  In order to calculate a threshold, we assume
that these parameters are equal across all four processes.  That is,
when we initialize a qubit to the $\ket{0}$ or $\ket{+}$ state, there
is a probability $p$ that the qubit is accidentally prepared into the
$\ket{1}$ or $\ket{-}$ state respectively.  Similarly, a readout error
in either the $X$ or the $Z$ basis will yield the incorrect result
with the same probability $p$.  A memory error is the application of
$X$, $Y$ or $Z$, each with probability $p/3$ to an idle qubit.  A
two-qubit gate error is the application of one of the 15 nontrivial
tensor products of $I$, $X$, $Y$ and $Z$, each with probability
$p/15$, after an ideal application of the two-qubit gate.  Each of
these processes are performed in one unit of time.

Our threshold calculation proceeds by assuming that we are given ahead
of time some quantum state encoded in a distance-$d$ code (either
toric or surface code).  The state begins in the simultaneous $+1$
eigenstate of all the stabilizers.  Given such a state, we can
determine the average time the state is sustained, using the described
error correction procedure, before a logical failure: at each step,
all the data qubits are time-evolved, and the syndrome extracted,
which permits the surface to be error corrected.  This cycle repeats
until logical failure ensues.  The simulations are performed by
tracing the propagation of errors through the computer, not the entire
quantum state.

For ideal syndrome extraction, instead of simulating time to failure
directly, we also count the number of $k$-error failures and determine
the logical error rate per timestep (the reciprocal of time to
failure), using

\begin{equation}
  p_L^{(d)}(p_0)
    = \sum_{k=0}^{Q} A_d(k) p^k (1-p)^{(Q-k)},
      \quad p = \frac 2 3 p_0
\end{equation}

Here $p_0$ is the physical error rate, $Q(d)$ is the number of data
qubits in the lattice, and $A_d(k)$ is the number of failure causing
configurations as a result of $k$ errors in the distance-$d$ code.
The factor of $\frac 2 3$ is due to our error model, as of the three
possible errors $X$, $Y$, $Z$ only two contain bit-flips which can
lead to logical-$X$ failures.  Calculating $A_d$ exactly is
computationally intractible, however we can approximate it by randomly
testing for failure only a fixed large number of the total number of
configurations, $Q \choose k$, giving a failure ratio $r_k$ and
allowing the approximation $A_d \approx {Q \choose k} r_k$.

\begin{figure}
\centering
\includegraphics[width=0.7\linewidth]{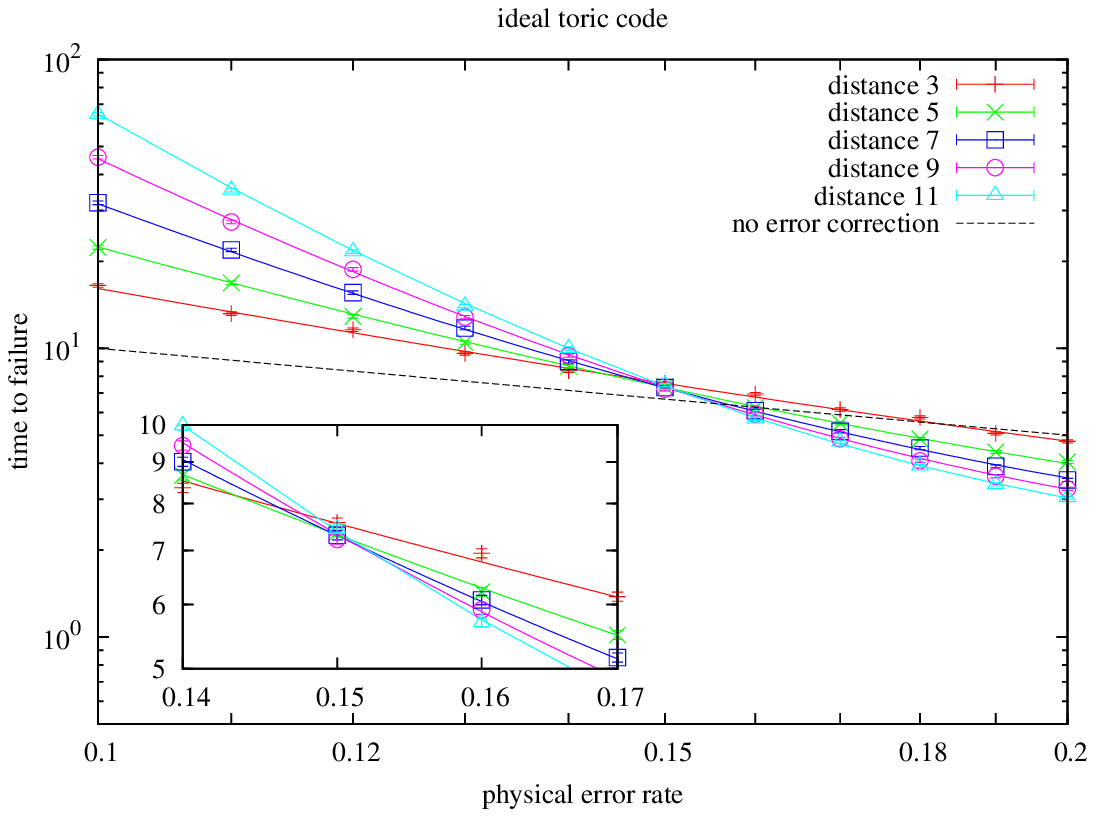}
\includegraphics[width=0.7\linewidth]{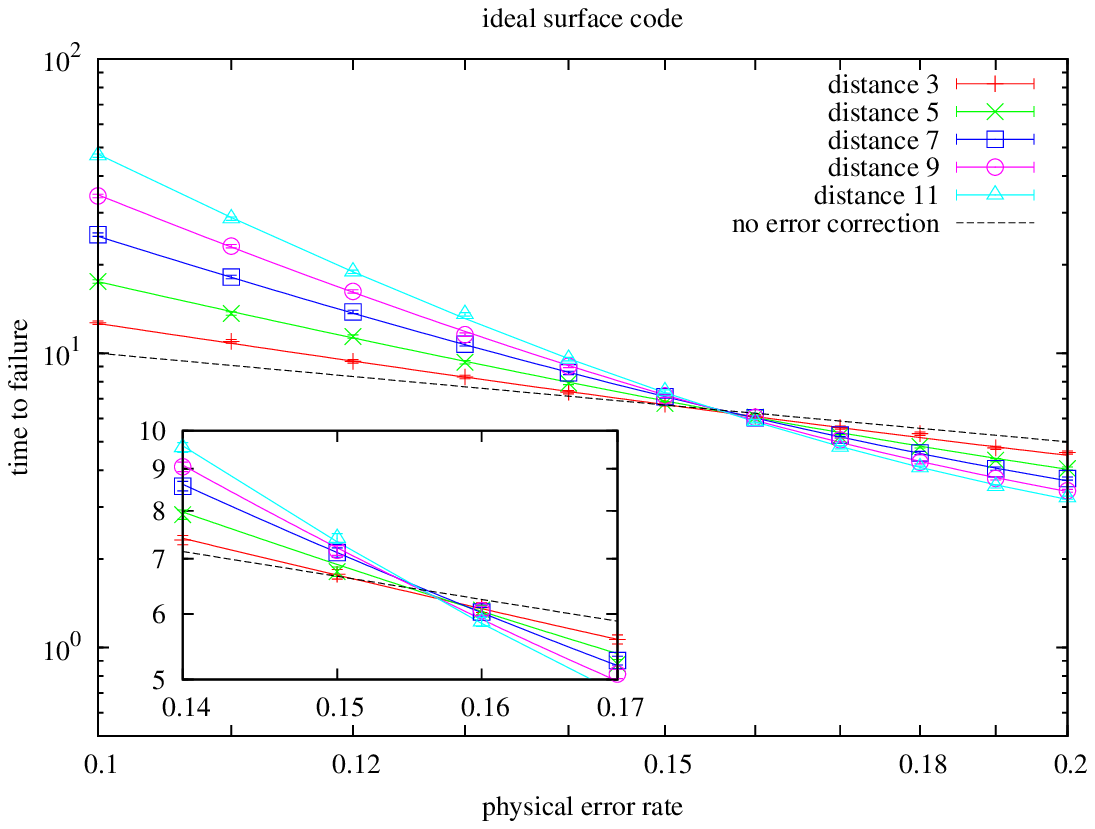}

\caption{Average life expectancy of a quantum state encoded in toric
and surface codes under error free syndrome extraction.  The error
bars represent the standard deviation in the value of the average.
The threshold is taken to be the error rate at the intersection
between successive distance codes, $p_{\mathrm{th}} = 0.155 \pm
0.005$.}
\label{fig:ideal result}
\end{figure}

The results for both the toric code and surface code under ideal
syndrome extraction are shown in figure \ref{fig:ideal result}.  The
data points are obtained directly from the time to failure
simulations, whereas the curves are from counting the number of
failures.  Both codes show threshold values of $p_{\mathrm{th}} =
0.155 \pm 0.005$.  Our value obtained by direct simulation of the
toric code is consistent with \cite{Denn02} which was obtained by
mapping the problem onto a two-dimensional random-bond Ising model and
identifying transitions between ordered phase and disordered phase.
Their value of $p = 11\%$ for $X$ (and $Z$) errors is recovered when
we account for the factor of $\frac 2 3$ due to our single qubit error
model.  Differences thereafter may be attributed to different
conventions when dangerous ambiguous syndromes arise.  

The non-ideal syndrome extraction case is a little trickier as we
assume that the final syndrome measurement is error free.  This is
resolved by assuming the current time $t$ is the final step and
extracting one more syndrome ideally.  Should correction fail using
this data, we record a failure time of $t$.  Otherwise we continue to
timestep $t+1$, recollecting the syndrome using a non-ideal syndrome
extraction cycle.

\begin{figure}
\centering
\includegraphics[width=0.7\linewidth]{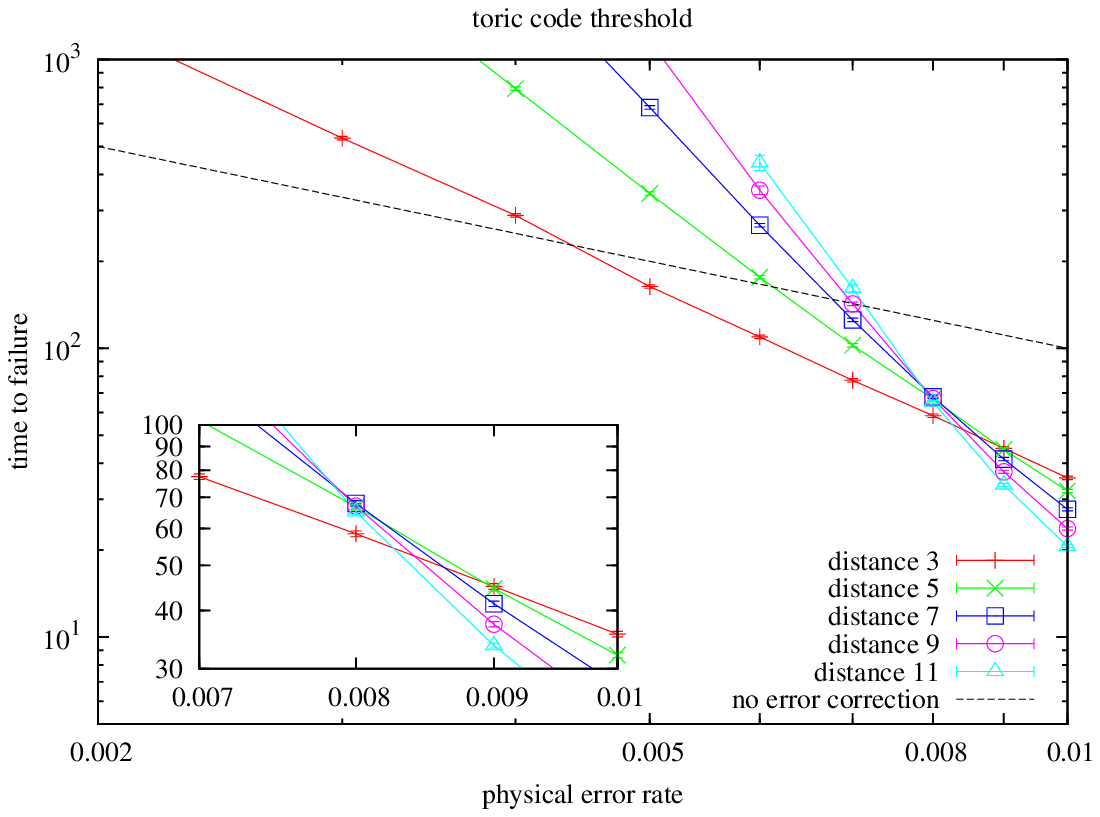}
\includegraphics[width=0.7\linewidth]{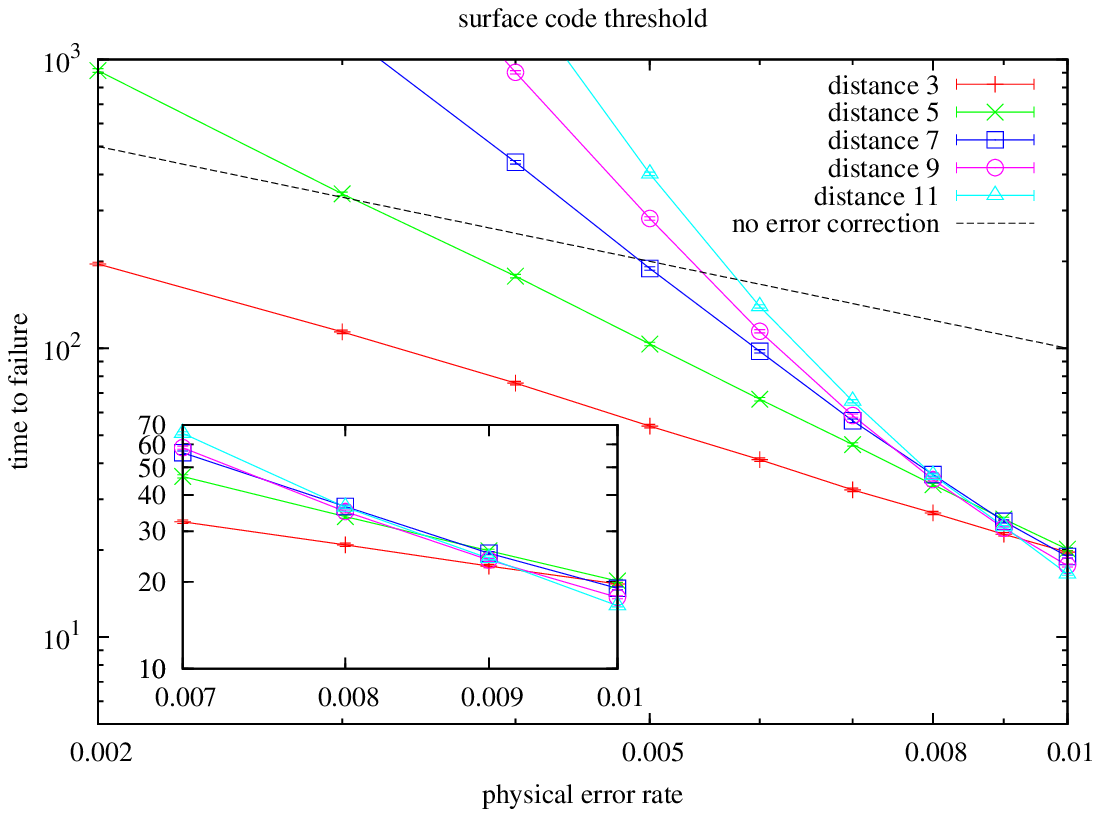}

\caption{Average life expectancy of a quantum state encoded in toric
and surface codes using syndrome extraction circuits of figure
\ref{fig:syndrome circuits}.  The toric code threshold is
$p_{\mathrm{th}} = 7.8 \times 10^{-3}$, and the surface code threshold
is expected to follow similarly.  Notice also that the surface code
now corrects one fewer error than the toric code for the same distance
code.}
\label{fig:faulty result}
\end{figure}

The results for the two codes now under non-ideal syndrome extraction
are shown in figure \ref{fig:faulty result}.  The toric code shows a
threshold error rate of $p_{\mathrm{th}} = 7.8 \times 10^{-3}$,
coinciding with previously known results \cite{Raus07}.  Slight
differences in value may be attributed to the different methodologies
of obtaining the threshold.

The surface code pseudo-thresholds fluctuate around the same region,
part of which can be put down to the result of boundary effects.
Because the boundary consists of only weight-$3$ stabilizers, syndrome
qubits on the boundary are more reliable than those in the bulk.  Thus
the effective error rate per timestep (syndrome extraction cycle) is
not uniform for different sized lattices; smaller lattices tend to be
more reliable as the ratio of boundary to bulk qubits is large.  By
this rationale, one would expect to observe surface code
pseudo-thresholds initially intersecting above the toric code
threshold and then monotonically converging towards it.  This
monotonic trend is not observed once we approach distance-$9$,
suggesting there is still more unexplained behaviour.  Nevertheless,
our results indicate that the asymptotic threshold for the surface
code threshold is in the vicinity of the toric code.

Another difference between the two codes is the maximum number of
errors that each can correct; a distance-$d$ toric code can fail with
$(d+1)/2$ errors, whereas the surface code can fail with only
$(d-1)/2$ errors.  The is due to correlated errors during the syndrome
extraction circuit.  One can find explicit examples where a single
error during syndrome extraction in the surface code appears na\"ively
as two errors, thus causing failure in a distance-$3$ code.  The toric
code, as a result of its wrap-around boundaries and hence somewhat
redundant extra row of syndrome qubits, avoids this behaviour.


\section{Conclusion}

We have described an efficient error correction scheme implementable
in practice.  Using this graphical approach for error correction, we
find the toric code threshold is $p_{\mathrm{th}} = 7.8 \times
10^{-3}$.  The surface code also shows a similar threshold.  Further
improvements to the graphical approach --- such as changing the
space-like and time-like edge weights to accurately reflect the
relative probabilities of data and readout errors, or considering more
error chains --- would boost the performance of these codes in real
life.

Error correction is usually thought of as performing this matching
algorithm only when necessary, for example when encountering
non-Clifford gates, after accumulating syndrome information over a
long stretch of time.  This delayed error correction means one must
build and match a large graph, which can potentially become a
bottleneck.  However, even without introducing approximations, the
graph can break into independent components thus the problem becomes
highly parallelisable.  Furthermore, it is possible to identify and
match many disconnected graph components before encountering these
non-Clifford gates, thus reducing the size of the task when one is
finally forced to error correct.

\nonumsection{Acknowledgements}

We thank Charles Hill, and Zachary Evans for their helpful
suggestions.
This work was supported by the Australian Research Council, the
Australian Government, and the US National Security Agency (NSA) and
the Army Research Office (ARO) under contract number W911NF-08-1-0527.

\nonumsection{References}
\bibliography{paper}
\bibliographystyle{unsrt}

\end{document}